\documentclass{moriond}
\usepackage{amssymb}
\usepackage{cancel}

\def\Journal#1#2#3#4{{#1} {\bf #2}, #3 (#4)}

\def\PLB{{\em Phys. Lett.}  B}
\def\PRL{\em Phys. Rev. Lett.}
\def\PRD{{\em Phys. Rev.} D}

\def\be{\begin{equation}}
\def\ee{\end{equation}}
\def\bea{\begin{eqnarray}}
\def\eea{\end{eqnarray}}

\newcommand{\Photo}{\includegraphics[height=35mm]{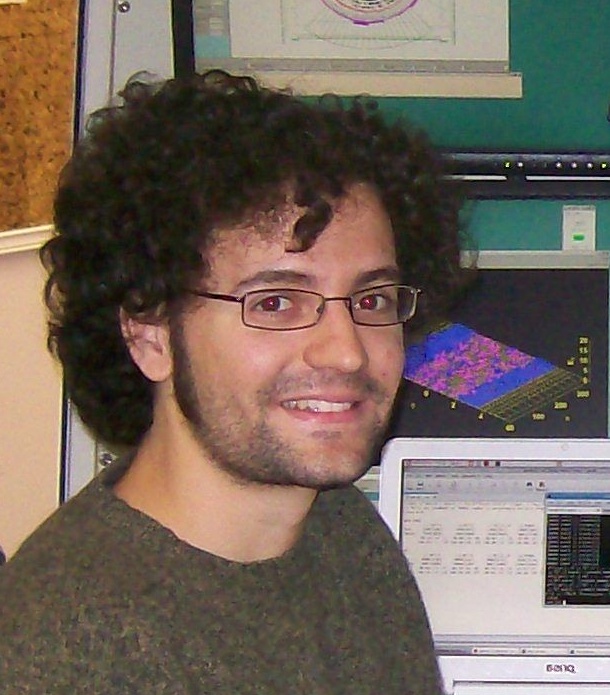}}

\begin{document}
\vspace*{4cm}
\title{TEVATRON MEASUREMENTS ON STANDARD MODEL HIGGS}

\author{Federico Sforza\\ {\small on behalf of the CDF and D0 Collaborations}}

\address{Max-Planck-Institut f\"ur Physik,\\
F\"ohringer Ring 6, M\"unchen, Germany}

\maketitle\abstracts{
We present the study of the SM Higgs properties obtained from the combined analysis of the up-to 10 fb$^{-1}$ dataset collected by the CDF and D0 experiments during the $p\bar{p}$ collision at $\sqrt{s}=1.96$~TeV of Tevatron Run II. The observed local significance for the SM Higgs boson signal is of 3.0$\sigma$ at $m_H=125$ GeV/c$^2$. After a brief review of analysis channels  contributing the most, where the Higgs boson decays to a pair of $W$ bosons or to a pair of $b$-quarks jets, the signal production cross section and its couplings to fermions and vector bosons are analyzed. Other presented results are the recent study of the spin and parity of the SM Higgs performed by the D0 collaboration, leading to 3$\sigma$ level expected exclusion of the JP$=0^{-}$ and JP$=2^{+}$ hypothesis, and the investigation of exotics final states with invisible decay products of the Higgs, excluded by the CDF collaboration for masses below 120 GeV.}

\section{Introduction}

The study of the up-to 10~fb$^{-1}$ datasets of $p\bar{p}$ collisions recorded by the CDF and D0 experiments during the Tevatron Run-II contributes significantly to the understanding of the properties of the SM Higgs particle discovered in 2012 by the ATLAS and CMS experiments at LHC\cite{h_LHC}. The CDF and D0 datasets have been collected at a center-of-mass energy of 1.96~TeV resulting in about a factor 20 lower Higgs signal production rate with respect to LHC one, however the advanced analysis techniques, often based on Multi-Variate Algorithms (MVA) signal discriminants, and the combined analysis of the datasets of both experiments, result in approximately a 2$\sigma$ sensitivity to the SM Higgs production, with $m_H=125$~GeV$/c^2$, at the Tevatron.

\section{Overview of SM Higgs Analysis Channels at the Tevatron}

The Tevatron experiments are sensitive to the SM Higgs production in a range of hypothetical mass between $90$ and $200$~GeV thanks to the combination of several analysis channels. 

For $m_H\gtrsim 135$~GeV, the gluon fusion ($gg\to H$) process, the dominant Higgs production mode with a cross section\cite{h_cx} of $\mathcal{O}(1)$~pb, can be investigated using the decay of the Higgs to a pair of $W$ bosons. Such decay is predicted to have a BR of approximately 30\% for $m_H=130$~GeV and rising at higher Higgs masses, therefore the clean di-lepton final state resulting from the $W\to \ell\nu$ decay of both the $W$s, is used to reject the overwhelming multi-jet background which has $\mathcal{O}(1)$~mb production cross section. Both the CDF and D0 collaborations investigated this channel\cite{WW} selecting candidates events with two charged leptons of opposite sign, a large imbalance in the total reconstructed transverse energy ($\cancel{E}_T$) signaling the presence of two undetected neutrinos, and zero or low jet activity. The mass of the final state, a very sensitive variable in the search for a resonance, can not be fully recontructed because of the presence of two neutrinos, therefore MVA have been used to combine relevant signal and background kinematic information into one variable which maximise the signal over background ($s/b$) discrimination. The sensitivity of the analysis is also improved by categorizing the candidate events according to the different $s/b$ resulting from lepton flavor (which include hadronic $\tau$ decay), lepton identification purity, and jet multiplicity of the final state.

For $m_H\lesssim 135$~GeV, the $VH$ associate production with the Higgs decaying to a pair of \mbox{$b$-quarks} ($VH\to b\bar{b}$) provides the highest sensitivity analysis channel because of the largest predicted SM branching fraction ($\mathcal{BR}(H\to b\bar{b})\approx 57\%$ for $m_H=125$~GeV/$c^2$), and the leptonic decay of the vector boson allows for feasible online selection and offline background rejection. The $VH\to b\bar{b}$ analysis channel is classified according to three specific signatures: $ZH\to\ell\ell+b\bar{b}$, $WH\to\ell + \cancel{E}_T +b\bar{b}$, and $W/Z H\to \cancel{E}_T +b\bar{b}$, where $\ell$ is an electron or a muon and the $\cancel{E}_T$ reveals a neutrino passing through the detector or a lepton which has not been identified. Each of the three signatures has been investigated by the CDF and D0 collaborations in independent and highly optimized analyses\cite{VH}, the main common features between them are: the selection of a final states enriched in $b-$jets by using multi-variate b-tagging algorithms able to reduce the $V+$jets background by a factor $10^2$, a still large fraction of irreducible backgrounds (mainly $t\bar{t}$ production and $V+$heavy flavor jets) estimated with a mixture of simulation and data-driven techniques, the final signal-to-background discrimination based on MVA giving a  $10-25$\%  sensitivity increase over the use of the di-jet invariant mass distribution. 

\begin{figure}
  \begin{minipage}{0.5\linewidth}
    \centerline{\includegraphics[width=0.99\textwidth]{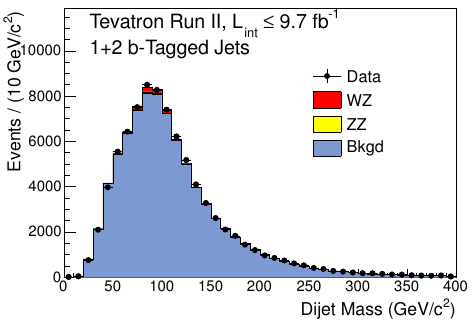}}
  \end{minipage}
  \hfill
  \begin{minipage}{0.5\linewidth}
    \centerline{\includegraphics[width=0.99\textwidth]{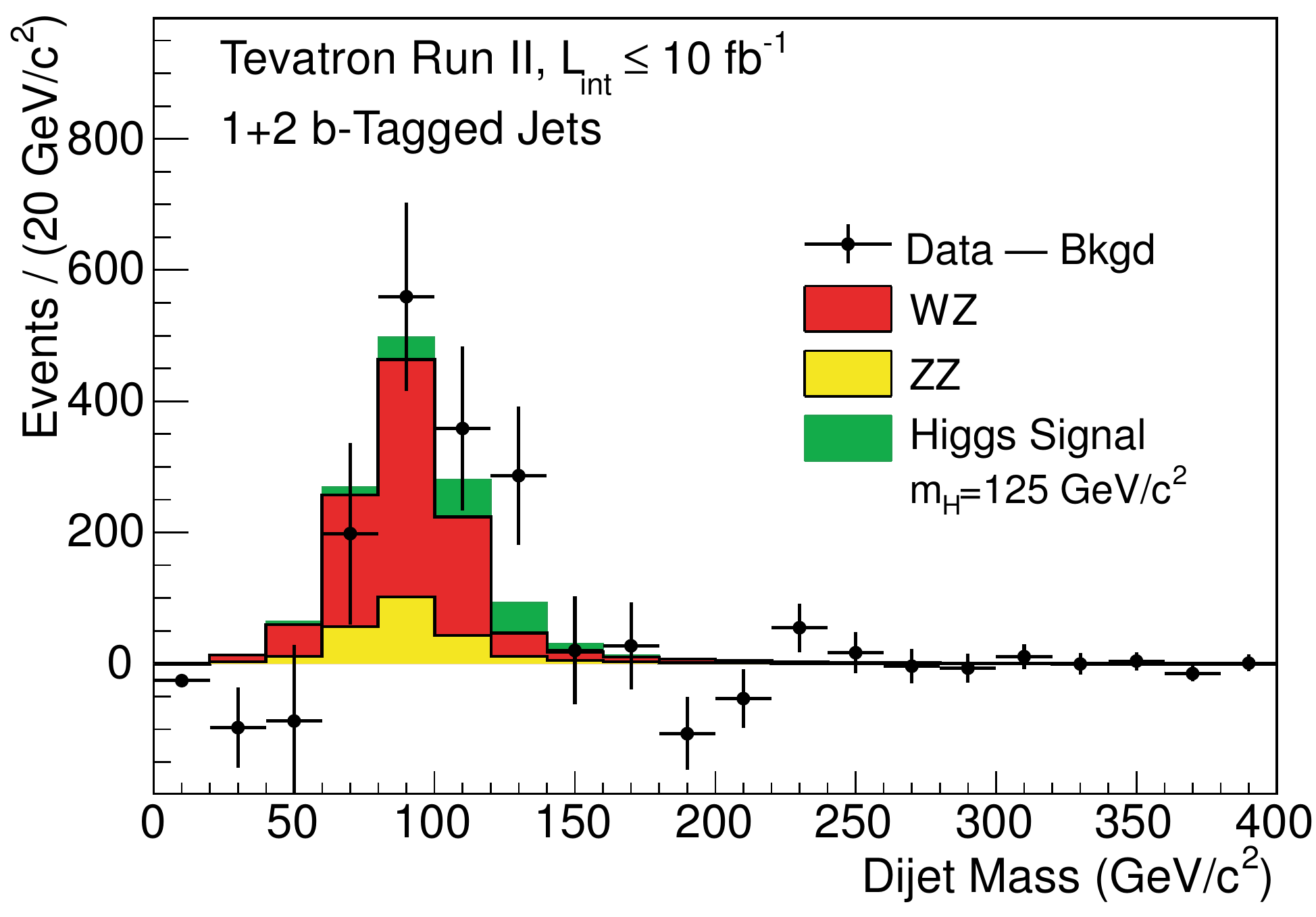}}    
  \end{minipage}
  \caption{Heavy flavor enriched di-jet mass spectrum of candidate events of the $VH\to b\bar{b}$ analysis channels after fit to the data withing systematic uncertainties. The estimated SM contributions are shown on the left plot while the background is subtracted on the right plot and the measured $Z$ and Higgs contributions are compared to data.}\label{fig:cmbb}
\end{figure}

A direct validation of these complex analyses is obtained by applying the same techniques to the search for a very similar signal: the $VZ$ associate production with the $Z$ decaying to a pair of Heavy Flavor (HF) quarks. With respect to the $VH\to b\bar{b}$ analysis with $m_H=125$~GeV the signal and the background yields are about 6 and 2 times higher respectively. The combination of the all the different search channels of both the CDF and D0 experiments, described in more detail in the next section, gives a strong evidence, clearly visible in the HF enriched di-jet mass spectrum of Fig.~\ref{fig:cmbb}, for the $VZ\to HF$ process: with a $p$-values of $4.6\sigma$ with respect to the background only hypothesis and a measured cross section of  $\sigma_{VZ}=3.0\pm0.6(stat.)\pm 0.7(syst.)$~pb, with a SM expectation of $\sigma_{VZ}=4.4\pm 0.3$~pb.

On the top of the main analysis channels, $gg\to H\to WW$ and $VH\to b\bar{b}$, a variety of searches with different final states ($H\to\gamma\gamma$, $H\to\tau\tau$, $H\to ZZ$, $t\bar{t}H$)  listed in\cite{Combo} provide a smaller increase in sensitivity across the SM Higgs mass range.

\section{SM Higgs Cross Section and Coupling Measurements from the Combination of CDF and D0 Analyses}

\begin{figure}[!h]
  \begin{minipage}{0.46\linewidth}
    \centerline{\includegraphics[width=0.99\textwidth]{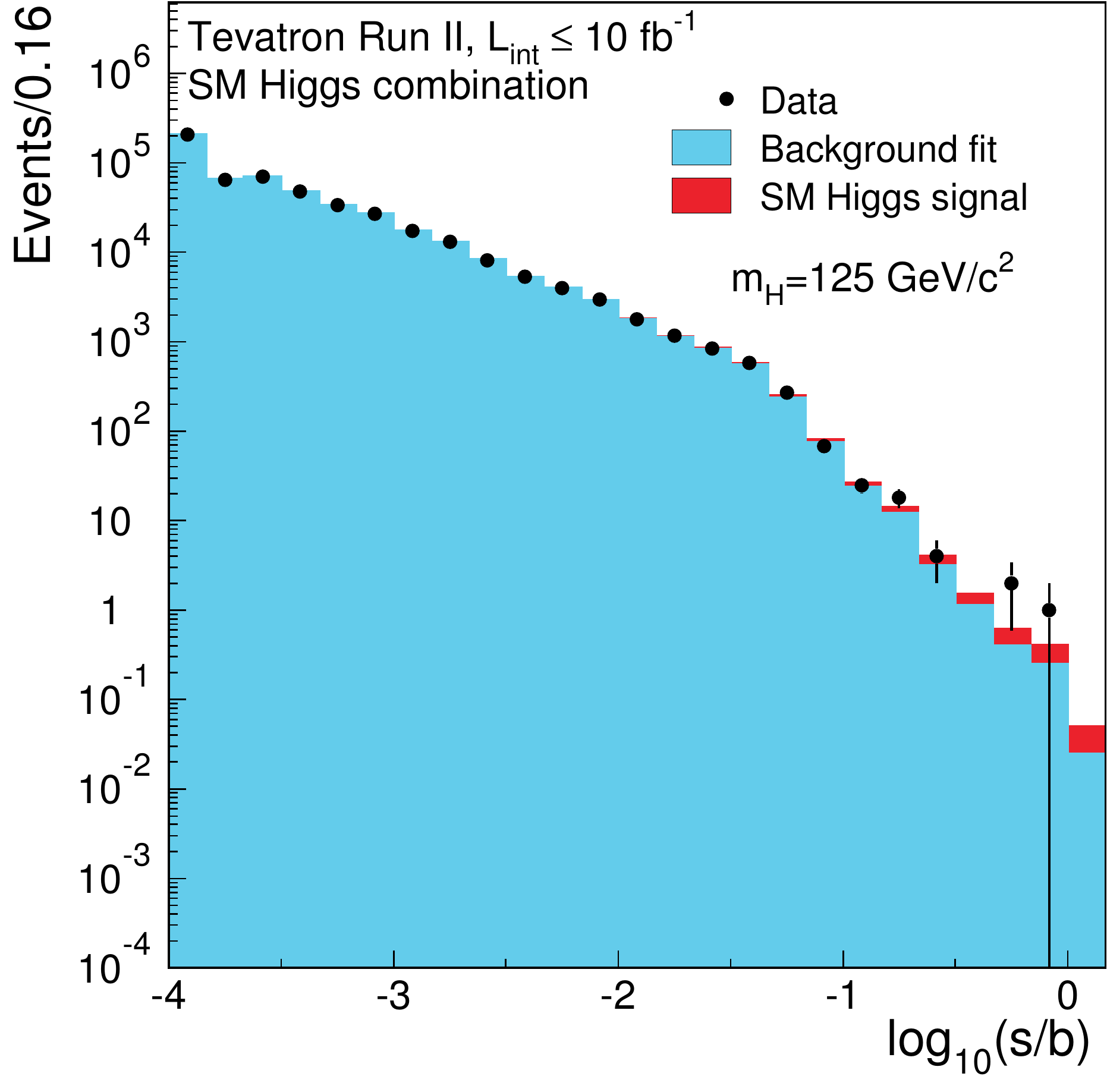}}
  \end{minipage}
  \hfill
  \begin{minipage}{0.54\linewidth}
    \centerline{\includegraphics[width=0.92\textwidth]{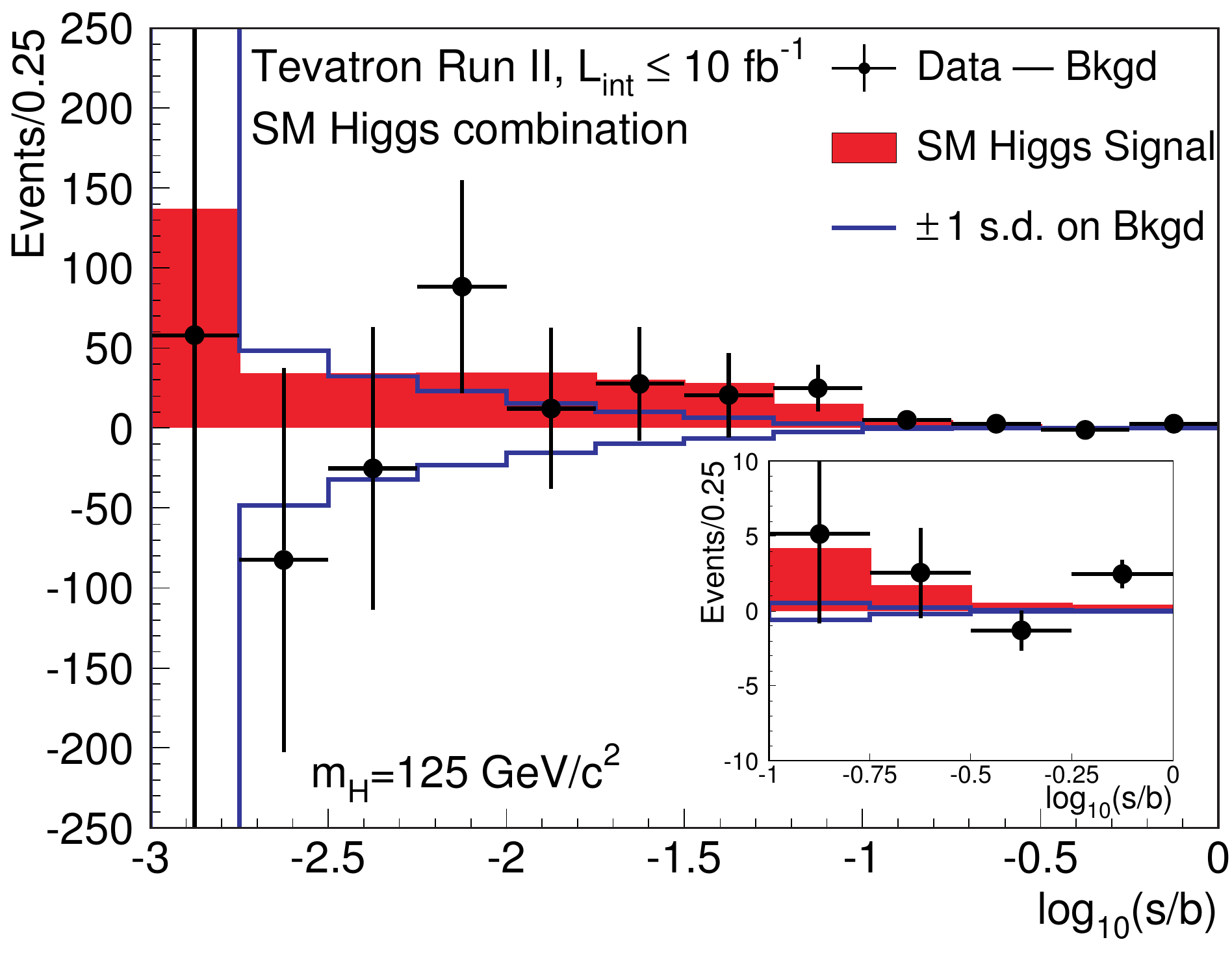}}    
  \end{minipage}
  \caption{Distribution of all the candidate events collected in the Tevatron SM Higgs analyses and ordered according the expected $\log_{10}(s/b)$. The plot on the left shows the data and the background agreement, after fit to data constraining the systematic uncertainties, the SM Higgs signal for $m_H=125$~GeV$/c^2$ is shown in red. The plot on the right shows the most significant bins after background subtraction.}\label{fig:comb_sob}
\end{figure}

As described in details in Ref.~\cite{Combo}, several properties of the SM Higgs boson can be constrained by the combined analysis of all the channels described in the previous section. Figure~\ref{fig:comb_sob} gives a global picture of such combination: all the candidate events are classified in bins of similar $s/b$ together with the corresponding background and signal contributions. After a fit to data withing the systematics uncertainties, the data-background agreement extends over five orders of magnitude, the highest $s/b$ bins show that the most signal-like events are consistent with the SM Higgs production with $m_H =125$~GeV$/c^2$.

The rigorous combination is done by building a global likelihood function from the product of the binned likelihood functions resulting from the data-background comparison in each of the discriminating variable obtained by the individual analysis channels.  The SM Higgs signal strength, $\mu=\sigma_{obs}/\sigma_{SM}$, is a free parameter of the likelihood, with branching fractions predicted by the SM, while systematics uncertainties are treated as nuisance parameters, with truncated Gaussian distributions, and they are often constrained by data of background dominated regions. Few of the systematics uncertainties are correlated between the CDF and D0 experiments (e.g. the uncertainties on  luminosity, theoretical cross sections, PDFs) while the experiment dependent ones are correlated across the analysis channels of each experiment (e.g. the uncertainties on lepton identification, b-tagging, jet energy scale, or specific ones on background estimate methodologies).

\begin{figure}[!h]
  \begin{minipage}{0.5\linewidth}
    \centerline{\includegraphics[width=0.99\textwidth,height=0.75\textwidth]{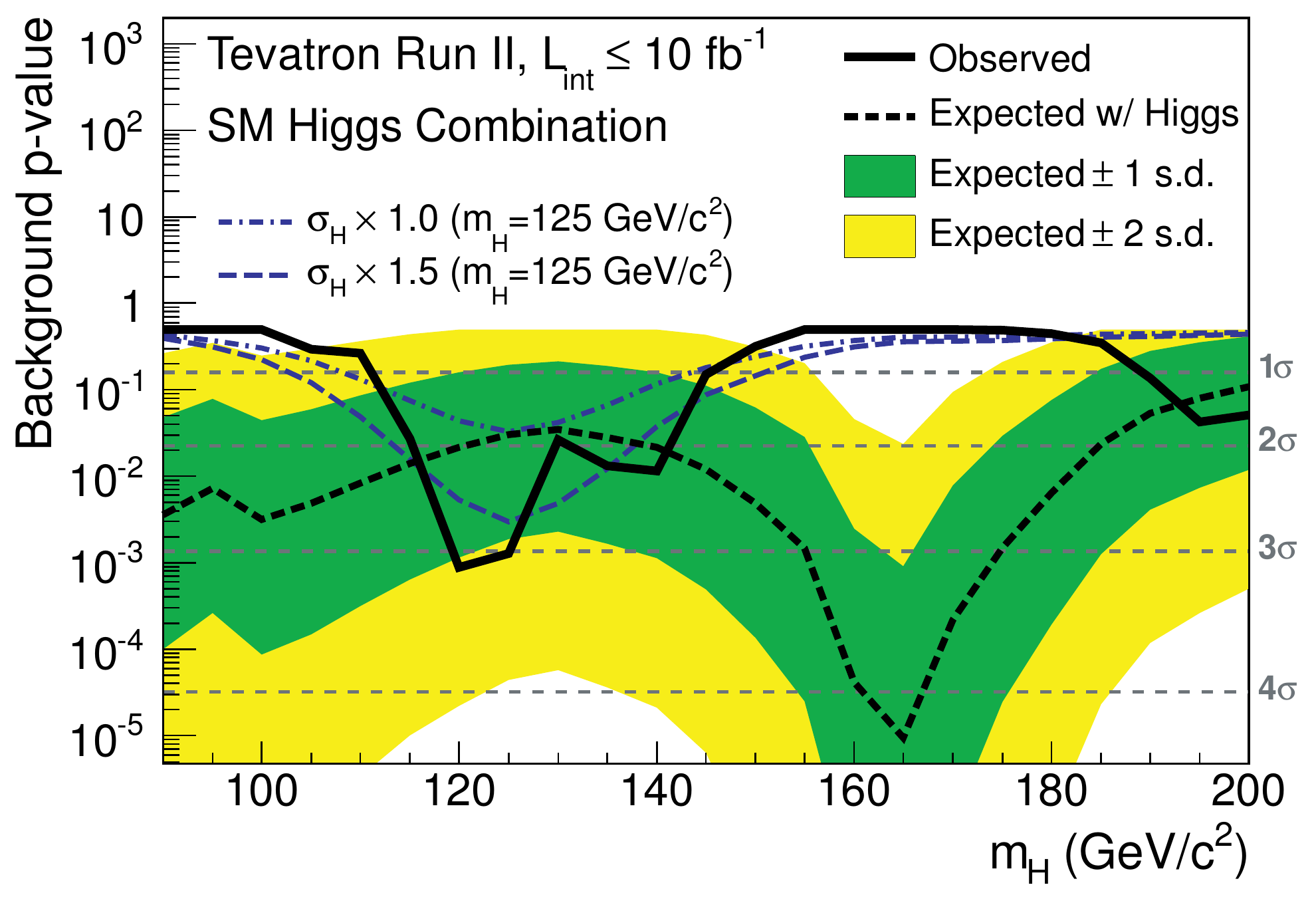}}
  \end{minipage}
\hfill
  \begin{minipage}{0.5\linewidth}
    \centerline{\includegraphics[width=0.99\textwidth,height=0.75\textwidth]{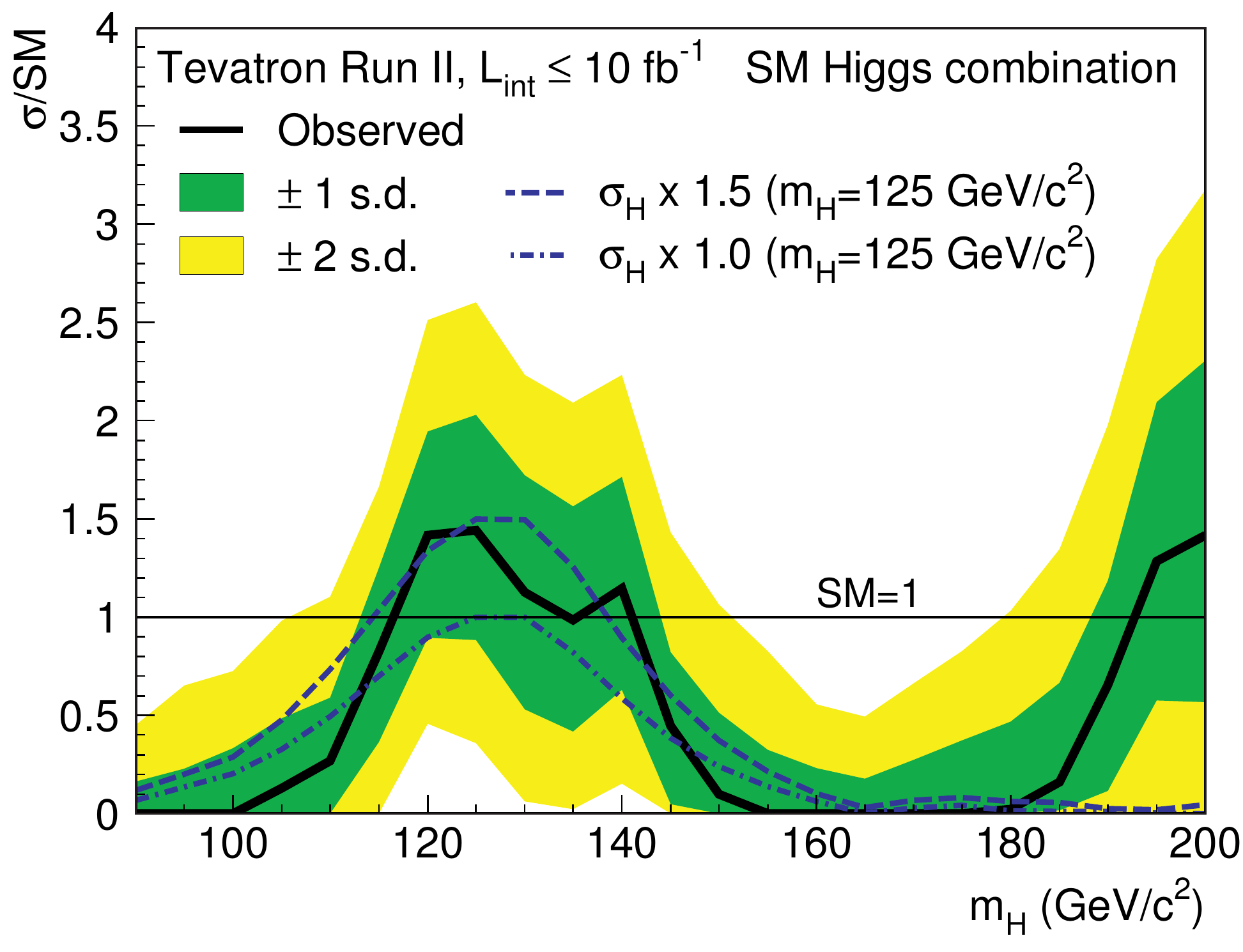}}
  \end{minipage}
  \caption{Results of the combination of all the Tevatron searches for a SM Higgs in the mass range  [90,200]~GeV/$c^2$: expected and observed $p-$value of the background only hypothesis (left), and the measured $\mu=\sigma_{obs}/\sigma_{SM}$ with respect to the SM expectation (right).}\label{fig:full_comb}
\end{figure}

The likelihood function is analyzed using both Bayesian and modified frequentest methodologies with consistent results. The outcome of the Bayesian analysis, chosen {\em a priori} to summarize the results, is shown in Figure~\ref{fig:full_comb}: the local $p-$value, evaluated against the background-only hypothesis, as a function of the Higgs boson mass shows a broad signal-like behaviour with a significance of 3.0$\sigma$ at $m_H=125$~GeV/$c^2$, for the same mass the observed signal strength is $\mu=1.44^{+0.59}_{ -− 0.56}$. Deviation from $\mu=0$ appears in the mass range $[115,145]$~GeV$/c^2$ because of the relatively low mass resolution of the sensitive analysis channels.

\begin{figure}[!h]
  \begin{minipage}{0.5\linewidth}
    \centerline{\includegraphics[width=0.99\textwidth, height=0.76\textwidth]{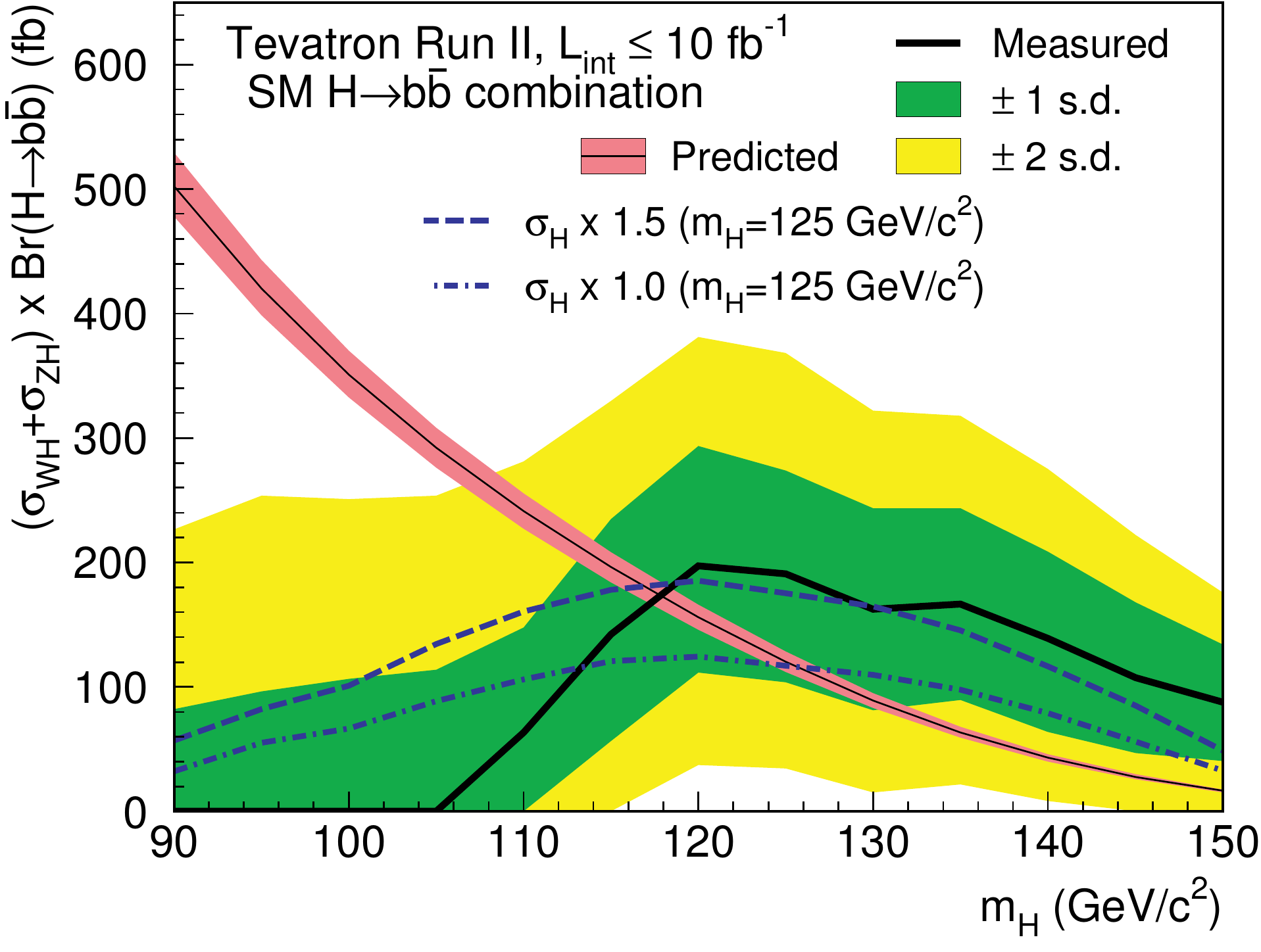}}
  \end{minipage}
\hfill
  \begin{minipage}{0.5\linewidth}
    \centerline{\includegraphics[width=0.99\textwidth,height=0.76\textwidth]{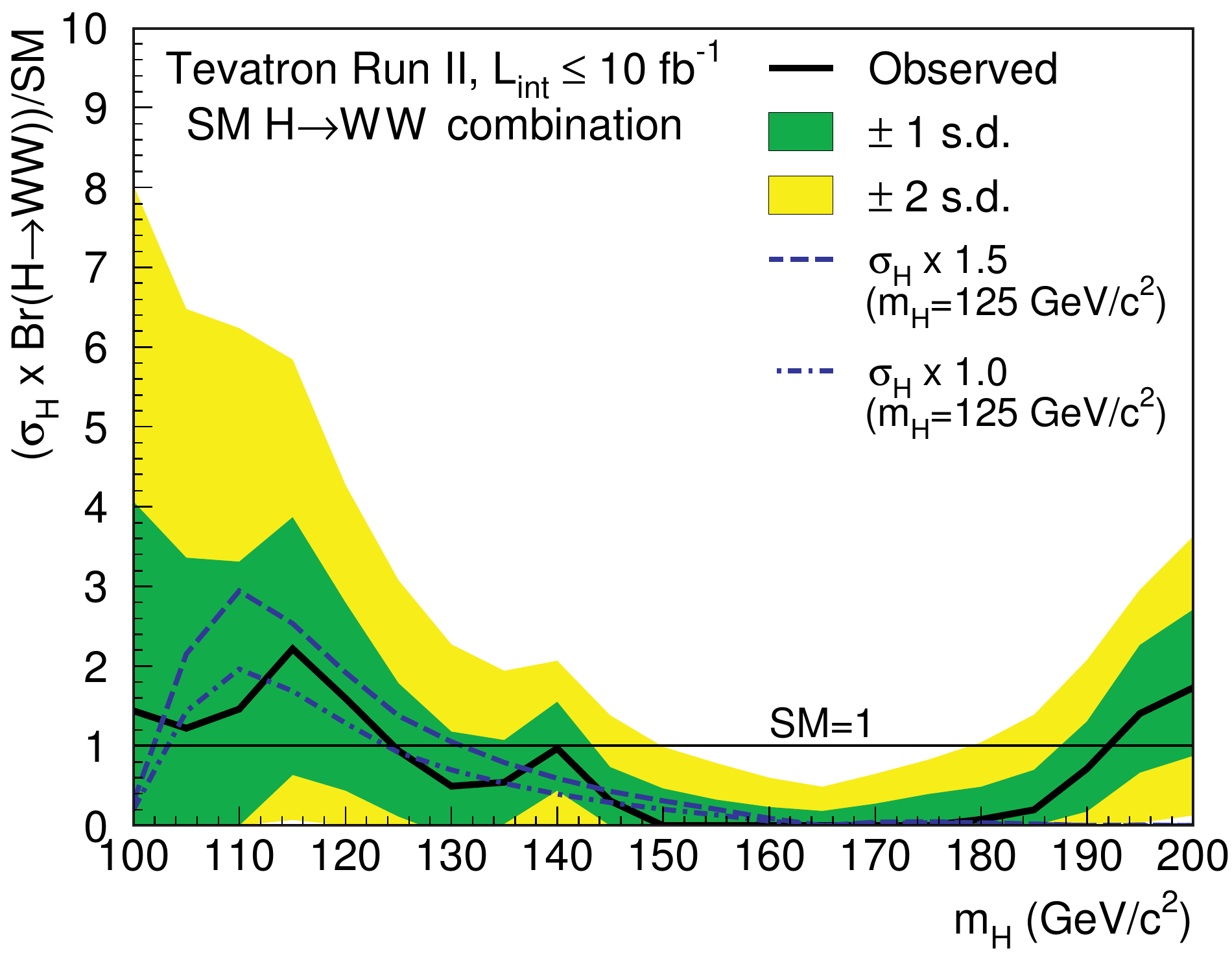}}
  \end{minipage}
  \caption{Measured cross section time branching fraction for the combination of the CDF and D0 analysis channels of the $VH\to b\bar{b}$ (left) and $H\to WW$ (right) processes as a function of the Higgs mass.}\label{fig:hchan_comb}
\end{figure}

Additional properties of the SM Higgs are extracted by analyzing the signal with respect to one decay channel at the time: $b\bar{b}$, $WW$, $\gamma\gamma$, and $\tau\tau$.  
The separate results are consistent with each other and with the presence of a SM Higgs with $m_H=125$~GeV$/c^2$, Fig.~\ref{fig:hchan_comb} shows the measured cross section times branching fraction as a function of the Higgs mass for the $H\to WW$ and $VH\to b\bar{b}$ channels. The analysis of the $VH\to b\bar{b}$ channels is particularly relevant not only because of the large weight in the Tevatron combined analysis but also because, with a best-fit measure of $\sigma_{VH\to b\bar{b}} = 0.19^{+0.08}_{-−0.09}$~pb, for $m_H = 125$~GeV$/c^2$, it is, at present, one of the most precise measurement of this process. 

\begin{figure}[!h]
  \begin{minipage}{0.5\linewidth}  
    \centerline{\includegraphics[width=0.99\textwidth]{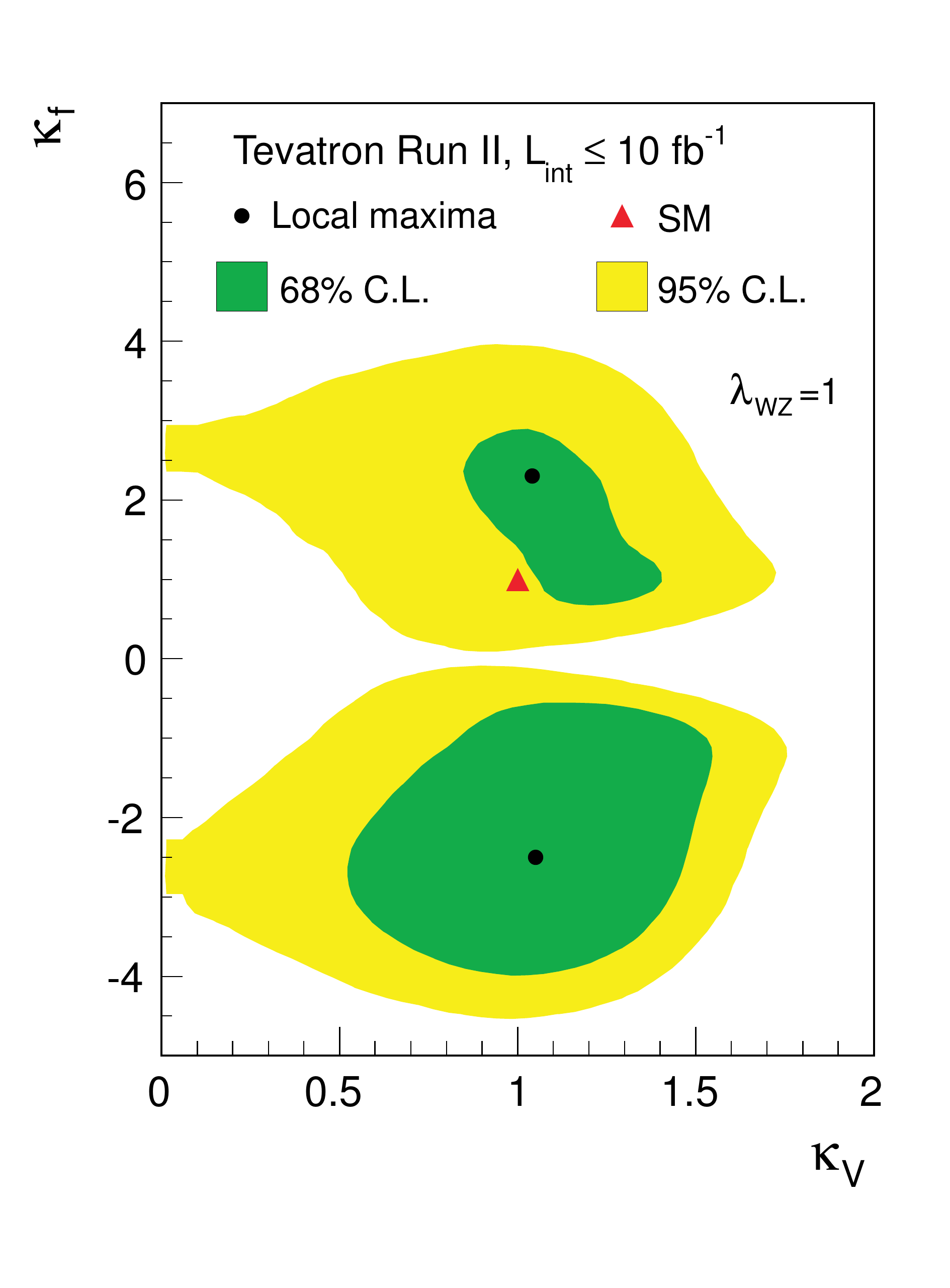}}
  \end{minipage}
\hfill
  \begin{minipage}{0.5\linewidth}
    \centerline{\includegraphics[width=0.99\textwidth]{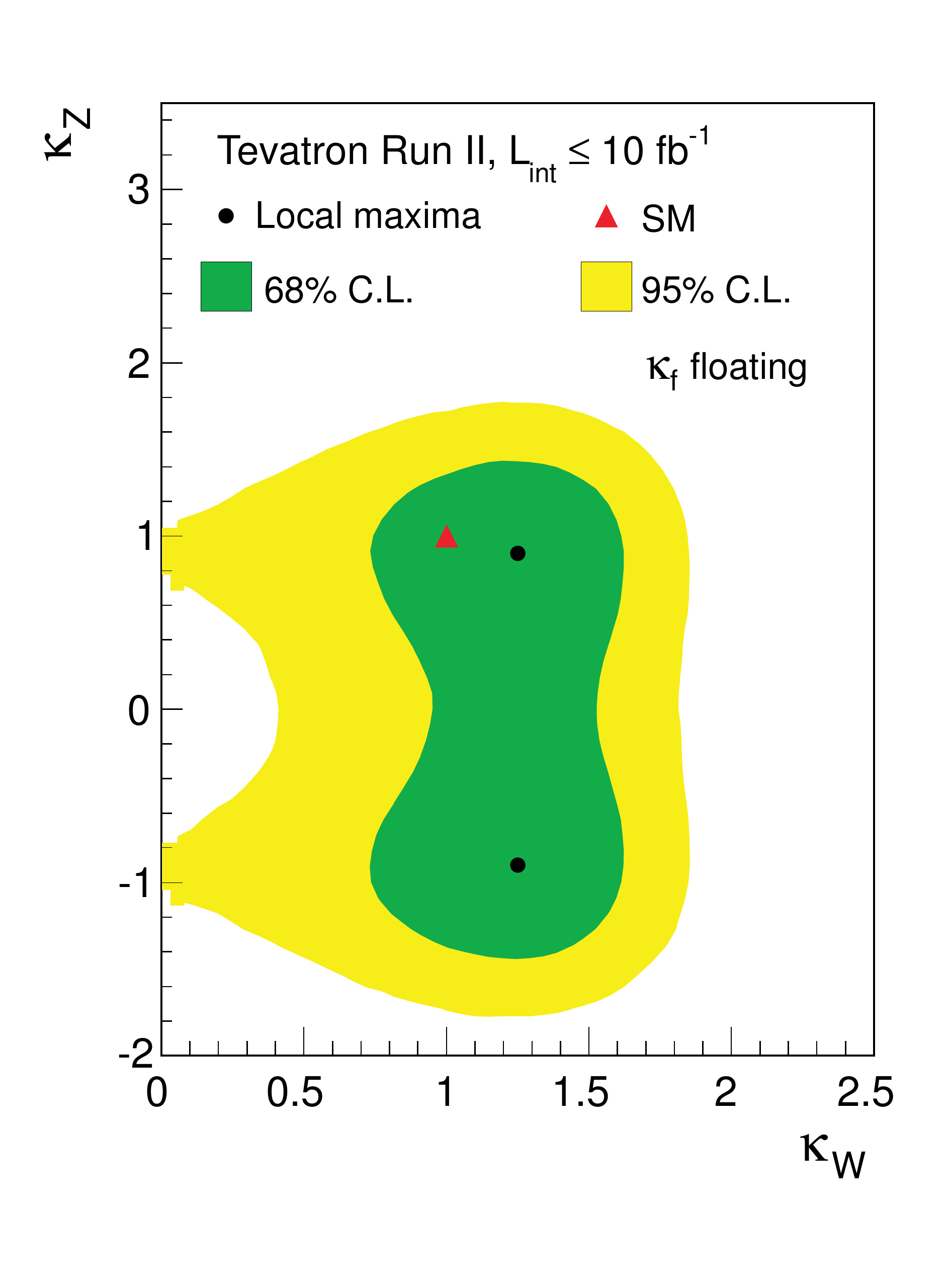}}
  \end{minipage}
  \caption{Two-dimensional Bayesian posterior confidence level distributions of the multiplicative scaling factors of the couplings of a SM Higgs, with $m_H=125$~GeV$/c^2$, to fermions ($\kappa_f$) {\em vs} vector bosons ($\kappa_V=\kappa_W=\kappa_Z$) on the left, and to the $W$ and $Z$ bosons separately ($\kappa_f$ is marginalized) on the right.}\label{fig:couplings}
\end{figure}

The channel-by-channel analysis  can also be used to study the SM couplings of the Higgs: as described in\cite{coupl_theo}, multiplicative scaling factors are associated to the coupling to fermions  ($\kappa_f$) and to $W/Z$ bosons ($\kappa_W$ and $\kappa_Z$),  their variation is then studied with Bayesian approach, assuming uniform prior.
Figure~\ref{fig:couplings} shows the two-dimensional confidence regions on the coupling parameters in case $\kappa_W = \kappa_Z = \kappa_V$, as predicted by the custodial symmetry, or in case the $\kappa_f$ coupling is marginalized in the likelihood so that $\kappa_W$ and $\kappa_z$ are studied separately.

\section{Analysis of Spin, Parity, and Exotic Final States}

Additional Higgs properties have been recently studied using the CDF or the D0 datasets: in particular the D0 collaboration investigated\cite{d0_spin} the spin and parity ($J^P$) of the Higgs particle while the CDF collaboration searched for Higgs boson decaying to invisible products\cite{cdf_inv}.

\begin{figure}[!h]
  \begin{minipage}{0.5\linewidth}
    \centerline{\includegraphics[width=0.99\textwidth]{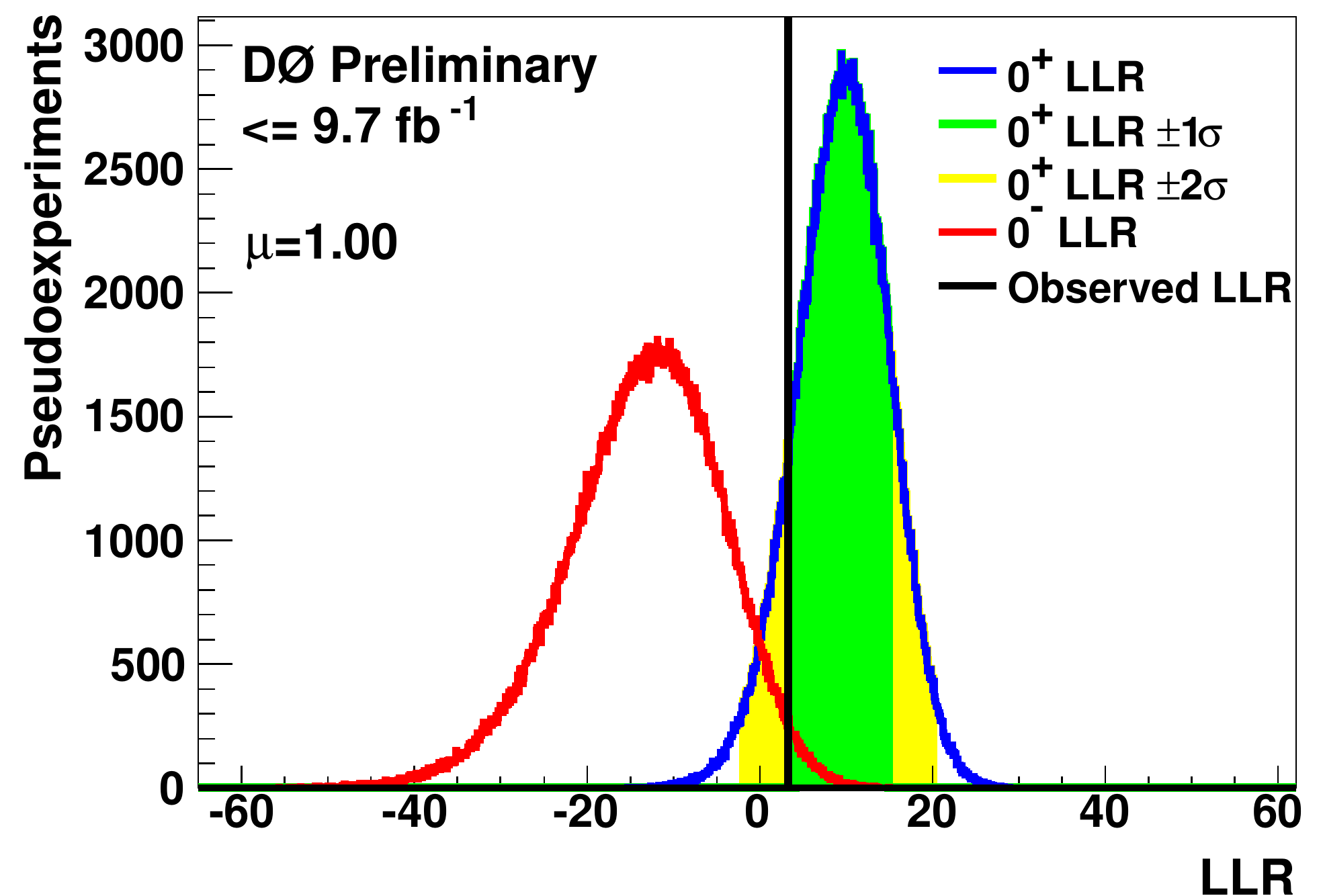}}
  \end{minipage}
\hfill
  \begin{minipage}{0.5\linewidth}
    \centerline{\includegraphics[width=0.99\textwidth]{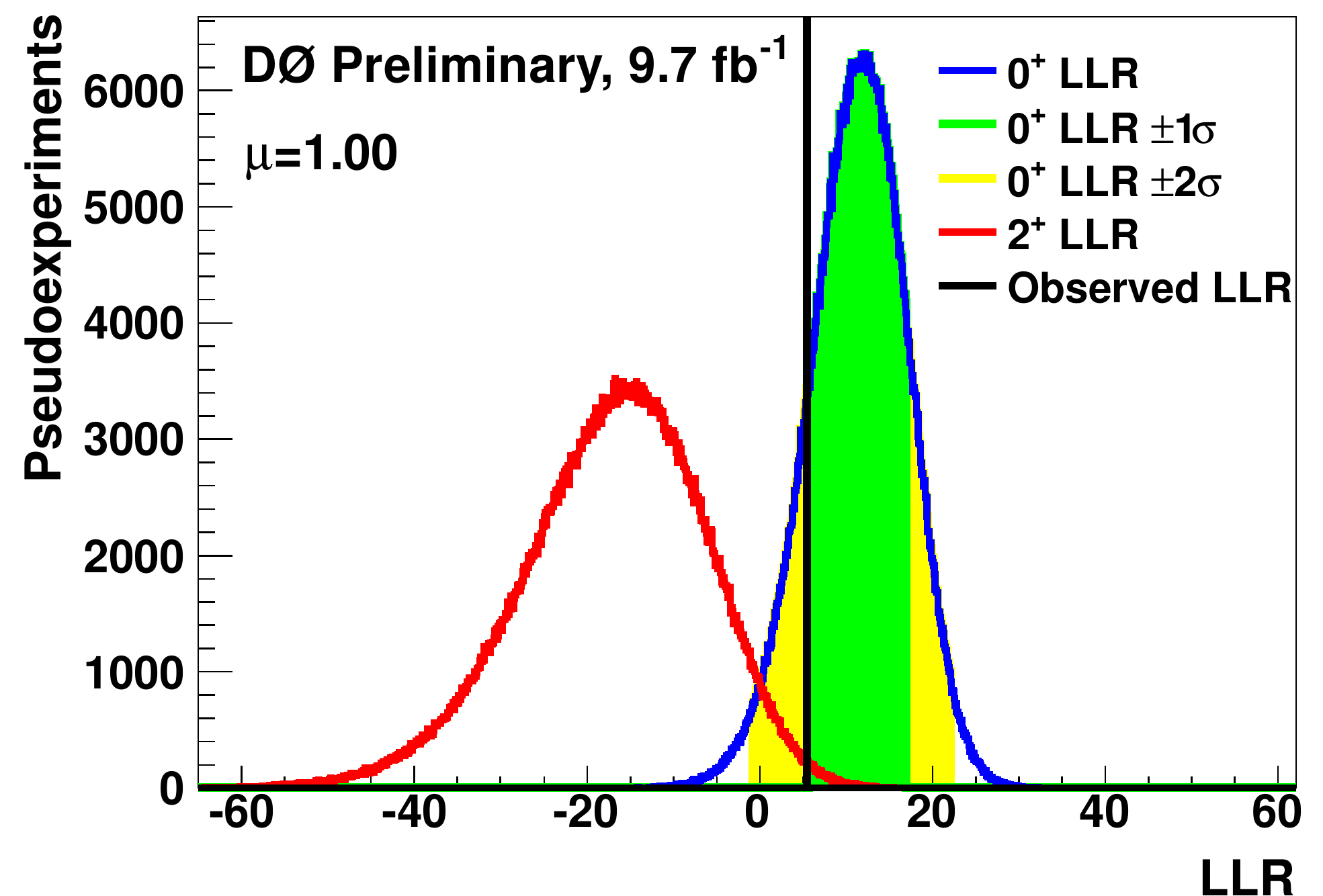}}
  \end{minipage}
  \caption{Results of the LLR for the data (black line), for the predicted background plus SM Higgs boson varied withing systematics (blue curve), and for the predicted background plus a Higgs particle with $J^P=0^{-}$ (red curve on the left plot) and $J^P=2^{+}$ (red curve on the right plot).}\label{fig:d0_spin}
\end{figure}

The analysis of the spin and parity of the Higgs is possible because some observables reconstructed in the $VH\to b\bar{b}$ channel, as the invariant mass of the $VH$ system\cite{spin_theo}, are sensitive to the $J^P$ of the Higgs-particle. Such study was performed by the D0 collaboration using the previously described $VH\to b\bar{b}$ analyses with no modification to the event selection but where sub-channels have been optimized according to the $s/b$ of the non-SM $J^P$ hypothesis, $J^P=0^-$ and  $J^P=2^+$. Assuming $m_H=125$~GeV/$c^2$ and the non-SM signal production rates to remain the same, the SM hypothesis $H_0$ is tested against the non-SM hypotheses $H_1$ by building, with pseudo-experiments, the distribution of the ratio between the maximised log likelihoods, $LLR= -2 \log \left( \mathcal{L}_{H_1}/ \mathcal{L}_{H_0}\right)$, and comparing it to the LLR obtained in data. Figure~\ref{fig:d0_spin} shows the LLR in case of the predicted SM production rates: the $J^P=0^-$ hypothesis is excluded with a $CL_s$ of 97.9\%, with an expected exclusion at the $3.1\sigma$ level, and the $J^P=2^+$ hypothesis is excluded with a $CL_s$ of 99.9\%, with an expected eclusion at the $3.2\sigma$ level.

\begin{figure}[!h]
 \begin{minipage}{0.5\linewidth}
    \centerline{\includegraphics[width=0.99\textwidth]{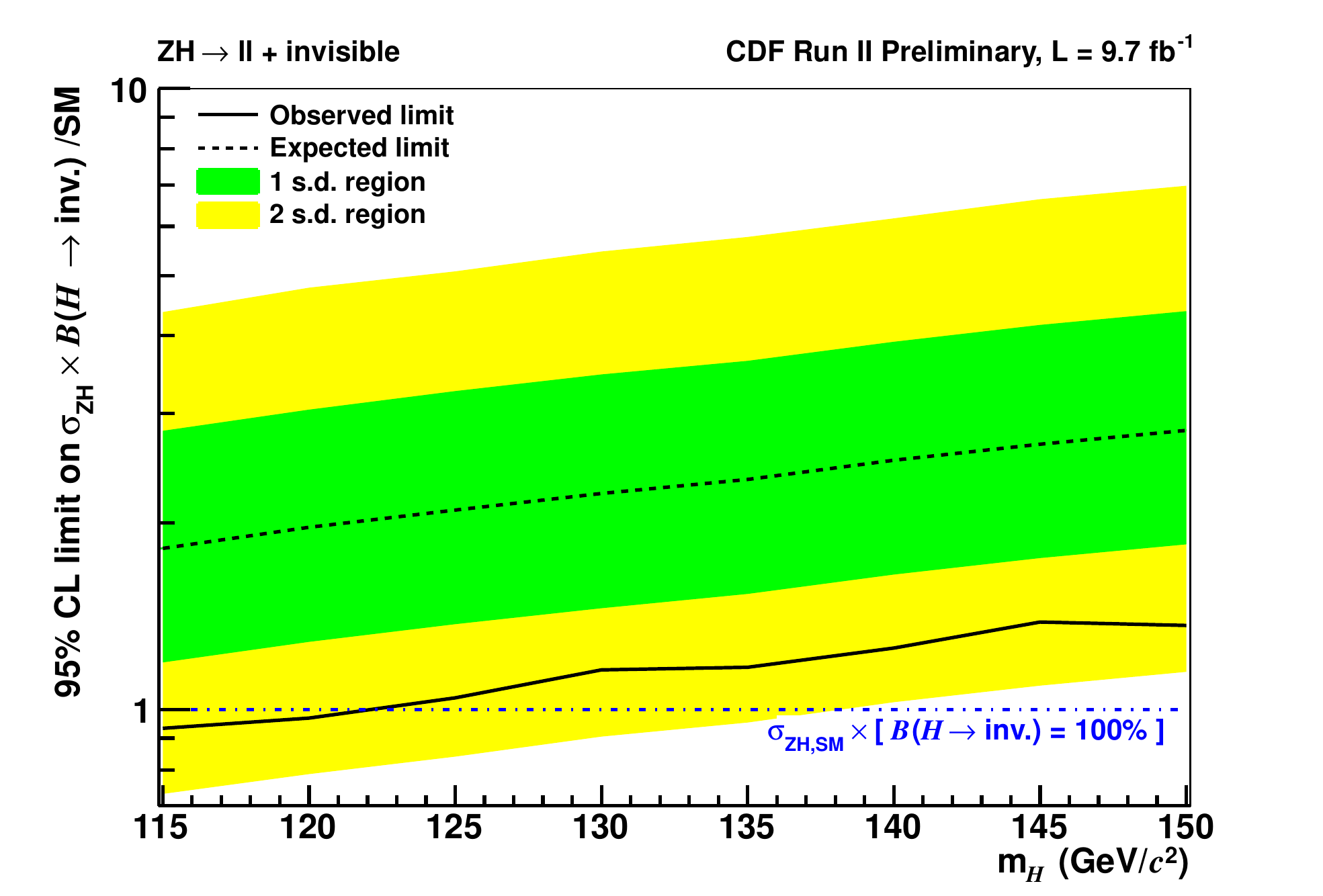}}
  \end{minipage}
\hfill
  \begin{minipage}{0.5\linewidth}
  \centerline{\includegraphics[width=0.99\textwidth]{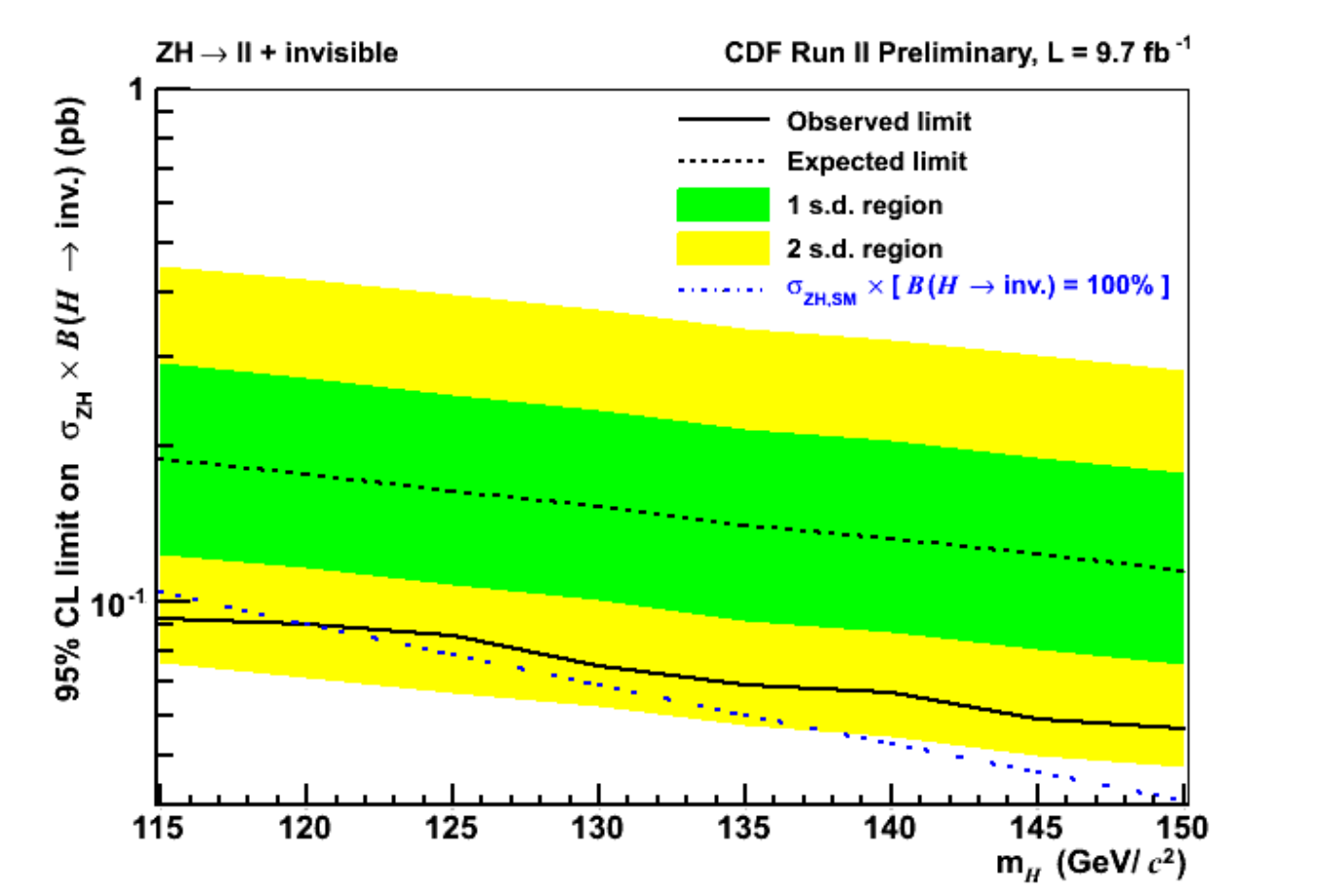}}
\end{minipage}
  \caption{The left part of the figure shows 95\% exclusion limits for a  Higgs decaying to invisible final states, in case $\mathcal{BR}(H\to\textrm{invisible})=100$\% is imposed.
The right part of the figure shows the cross section upper limits at 95\% confidence level for invisible decay of the Higgs boson when the  $\mathcal{BR}(H\to\textrm{invisible})=100$\% hypothesis is removed. Plots are shown for a Higgs boson produced in association with a $Z$ boson and a Higgs mass varied between  $115$ and $150$~GeV$/c^2$. }\label{fig:h_inv}
\end{figure}

The CDF collaboration studied a decay channel where the Higgs, produced in association with a leptonically decaying $Z$ boson, decays to un-detected particles ($H\to$invisible). 
The signal presence is investigated with a Bayesian methodology by building a likelihood function out of the distribution of the $\Delta R$ between the leptons in the background-only or in the background-plus-signal hypothesis, with the signal generated for $115<M_H<150$~GeV$/c^2$. Figure~\ref{fig:h_inv} shows, on the left part, the observed mass exclusion limits in the hypothesis that the signal is normalized to have $\mathcal{BR}(H\to\textrm{invisible})=100$\%, while, on the right part, such   hypothesis is removed allowing the derivation cross section upper limits. Masses of the Higgs below 120~GeV/$c^2$ are excluded at 95\% confidence level if  $\mathcal{BR}(H\to\textrm{invisible})=100$\% is imposed on the signal, while, removing such constraint, a production cross section upper limit of 90~fb is obtained for a Higgs boson with $M_H=125$~GeV$/c^2$.
  
\section{Conclusions}

Despite the end of the Tevatron data taking in 2011, the collected data continues to be relevant for the study of SM Higgs boson properties.
The combined analysis of the Tevatron Run II datasets collected by the CDF and D0 experiments show sensitivity to the SM Higgs signal in a mass range of $[90,200]$~GeV/$c^2$ with an excess, over the background only hypothesis, of $3\sigma$ significance at a Higgs mass of $125$~GeV/$c^2$. The dataset has been used to measure a total SM Higgs production cross section, with respect to the SM expectation, of $\sigma_{obs}/\sigma_{SM}=1.44^{+0.59}_{-0.56}$, the channel dependent cross-sections times branching fractions, and the couplings of the Higgs particle with fermions and vector bosons. The result of $\sigma_{VH\to b\bar{b}} = 0.19^{+0.08}_{-−0.09}$~pb has still at present comparable sensitivity to the LHC experiments.

Additional properties have been investigated recently by the CDF or by the D0 collaboration. In particular the D0 collaboration excluded, with a model dependent analysis, the $J^P=0^-$ and $J^P=2^+$ hypothesis at more than 97\% CL, with expectations above $3\sigma$. The CDF collaboration excluded at 95\% CL the possibility of invisible decay of the Higgs, produced in association with a $Z$ boson, for masses below 120~GeV$/c^2$, if $\mathcal{BR}(H\to\textrm{invisible})=100$\% is imposed, or for a cross section above 90~fb, if the $\mathcal{BR}(H\to\textrm{invisible})=100$\% hypothesis is removed and the Higgs mass is of  $125$~GeV$/^2$.

\section*{Acknowledgments}
I would like to thank the CDF and D0 collaboration for the opportunity to present these interesting results, the Max-Planck-Institut f\"ur Physik for its important support,
and the conference organizers for the beautiful and enriching experience of participating to the Electroweak session of the $49^{th}$ Rencontres de Moriond.

\end{document}